\documentclass[a4paper,11pt]{article}

\usepackage{amsmath,amssymb,amsthm,bbm,bm}  
\usepackage{authblk}  
\usepackage[british]{babel}
\usepackage{balance}
\usepackage[sorting=none,doi=false,isbn=false,giveninits=true,style=nature,url=false]{biblatex}           
\usepackage{booktabs} 
\usepackage[font=small]{caption}
\usepackage{color}
\usepackage{comment}
\usepackage{enumitem}
\usepackage[T1]{fontenc}
\usepackage[top=2.5cm, bottom=2.5cm, left=2.5cm, right=2.5cm]{geometry}
\usepackage{graphicx}
\usepackage{graphbox}       
\PassOptionsToPackage{hyphens}{url}  
\usepackage[colorlinks = true,
            linkcolor = OliveGreen,
            urlcolor  = blue,
            citecolor = MidnightBlue,
            anchorcolor = blue]{hyperref}
\usepackage{cleveref}  
\usepackage[utf8]{inputenc}
\usepackage{csquotes}  
\usepackage{libertine}  
\usepackage{mathtools}
\usepackage{microtype}      
\usepackage{nicefrac}
\usepackage{physics}
\usepackage{subfigure}
\usepackage[usenames,dvipsnames]{xcolor}

\newcommand{\dropcap}[1]{#1}  
\newcommand{\methods}{\emph{Methods}}  
\DeclareMathOperator{\Var}{Var}
\DeclareMathOperator{\pmse}{pmse}
\bibliography{references}

\title{Data-driven emergence of \\ convolutional structure in neural networks}

\author[1]{Alessandro Ingrosso\thanks{ingrosso@ictp.it}}
\author[2]{Sebastian Goldt\thanks{sgoldt@sissa.it}}

\affil[1]{The Abdus Salam International Centre for Theoretical Physics (ICTP), Strada Costiera 11, 34151 Trieste, Italy}
\affil[2]{International School of Advanced Studies (SISSA), via Bonomea 265, 34136 Trieste, Italy}

\date{\today}

\begin{document}

\maketitle

\begin{abstract}
  Exploiting data invariances is crucial for efficient learning
  in both artificial and biological neural circuits. Understanding how neural networks can discover appropriate representations capable of harnessing the underlying symmetries of their inputs is thus crucial in machine learning
  and neuroscience. Convolutional neural networks, for example, were designed to exploit translation symmetry and their capabilities triggered the first wave of deep learning successes. However, learning convolutions directly from translation-invariant data with a fully-connected network has so far proven elusive. Here, we show how
  initially fully-connected neural networks solving a discrimination task can learn a convolutional structure directly from their inputs, resulting in
  localised, space-tiling receptive fields. These receptive fields match the filters of a convolutional network trained on the same task. By carefully designing data models for the visual scene, we show that the emergence of this
  pattern is triggered by the non-Gaussian, higher-order local structure of the
  inputs, which has long been recognised as the hallmark of natural images.  We
  provide an analytical and numerical characterisation of the pattern-formation
  mechanism responsible for this phenomenon in a simple model and find
  an unexpected link between receptive field formation and tensor
  decomposition of higher-order input correlations. These results provide a new
  perspective on the development of low-level feature detectors in various
  sensory modalities, and pave the way for studying the impact of higher-order
  statistics on learning in neural networks.
\end{abstract}

\section*{Introduction}
\addcontentsline{toc}{section}{Introduction}

\dropcap{E}xploiting invariances in data is crucial for neural networks to learn efficient representations and to make accurate predictions. Translation invariance is a key symmetry in image processing, and lies at the heart of feed-forward~\cite{DiCarloZoccolanVisual, YaminsPerformanceOptimized} and recurrent~\cite{DiCarloFastRecurrent, NicoRecurrentBetter} models of the visual system. In the early sensory stage, the feature maps obtained by convolving a set of filters with an input arise from the collective action of localised receptive fields organised in a tessellation pattern.
The importance of receptive fields for understanding neural networks was recognised in the seminal work of Hubel and Wiesel on the early stages of the visual system~\cite{Hubel62}. Receptive fields remain a key building block in theoretical neuroscience~\cite{KnudsenOwl, Jones1987TheTS, DiCarloSomatosensory}, from the statistical formulation of single-neuron encoding~\cite{PaninskiStatistical, VidneModeling} to hierarchical models of cortical processing in various sensory modalities~\cite{HierarchicalPoggio,TheunissenNeuralProcessing}.
A key question in neuroscience is how these receptive fields are developed and what mechanism drives their spatial organisation.
The computational inquiry into how receptive fields can originate from image statistics goes back to the seminal work of Olshausen and Fields~\cite{olshausen1996emergence}, who showed that a specific unsupervised learning algorithm maximising sparseness of neural activity was sufficient for developing localised receptive fields, similar to those found in primary visual cortex.

In machine learning, convolutional neural networks~(CNNs)~\cite{lecun1990handwritten} were inspired by the ideas of Hubel and Wiesel~\cite{Hubel62} and rely on linear convolutions, followed by non-linear functions and pooling operations~\cite{lecun2015deep} that encourage translation invariance of the network output~\cite{scherer2010evaluation,schmidhuber2015deep,goodfellow2016deep}. CNNs classify images significantly better than vanilla, fully-connected (FC) networks, which do not take this symmetry explicitly into account~\cite{urban2017deep}. Since their success in computer vision~\cite{krizhevsky2012imagenet,lecun2015deep,simonyan2015very, he2016deep}, deep CNNs have served as a prime example for how encoding prior knowledge about data invariances into the network architecture can improve both sample- and
parameter-efficiency of learning.

Subsequent work has since engineered architectures and
representations capable of dealing with data characterised by different invariances and geometries, such as social or gene regulatory networks~\cite{CohenGroupEquivariant, CohenGaugeEquivariant,KondorCompact,rezende2019equivariant, CohenSpherical, BronsteinGeometric2017, SteerableCNN, BronsteinGeometric, Bronstein2021geometric}.
These invariances, however, are not always known beforehand. Deep scattering networks~\cite{mallat2012group, bruna2013invariant, mallat2016understanding} have been proposed as architectures that are invariant to a rich class of transformations. Another approach altogether would be to learn low-level feature detectors that take basic symmetries into account directly from data. In the case of images, the question thus becomes: \emph{can we learn convolutions from scratch?}

The hallmarks of convolutional structure that we are looking for are local connectivity, resulting in localised receptive fields, and the sharing of weights between neurons. Furthermore, we require that the local filters have to be applied across the whole image, \emph{i.e.}~the filters have to tile the sensory space. Uniform tiling of sensory space is crucial in our understanding of input processing in biological circuits, and a number of theoretical justifications have been given in terms of coding efficiency~\cite{DoiEfficientCoding, KarklinEfficientCoding}.

Fully-connected layers are expressive enough to implement such convolutional structure, with weights that are sparse (due to locality) and redundant (due to weight sharing). The emergence of localised receptive fields has been recently shown in unsupervised models such as autoencoders~\cite{benna2021place, FarrellAutoencoder} and Restricted Boltzmann Machines~\cite{harsh2020placecell}, or with the use of similarity-preserving learning rules~\cite{sengupta2018manifoldtiling}. However, learning convolutions directly
from data by training an initially fully-connected network on a discriminative
task has so far proven elusive: FC networks do not develop any of the hallmarks of convolutions without tailor-made regularisation techniques, and they perform significantly worse than convolutional networks~\cite{urban2017deep, dascoliNeedle, NeyshaburScratch}.
The problem thus lies in the learnability of the convolutional structure through the standard paradigm of machine learning (optimization of a cost function via first-order methods).

Here, we show that fully-connected neural networks can indeed learn a convolutional structure directly from their inputs if trained on data with non-Gaussian, higher-order local structure. We design a supervised classification task that fulfils these criteria, and show that the higher-order statistics of the inputs can drive the emergence of localised, space-tiling receptive fields.

\section*{Results}
\addcontentsline{toc}{section}{Results}

\phantomsection
\subsection*{Fully-connected networks can learn localised receptive fields from scratch}
\addcontentsline{toc}{subsection}{Fully-connected networks can learn localised receptive fields from scratch}

\begin{figure*}
    \centering
    \includegraphics[width=\textwidth]{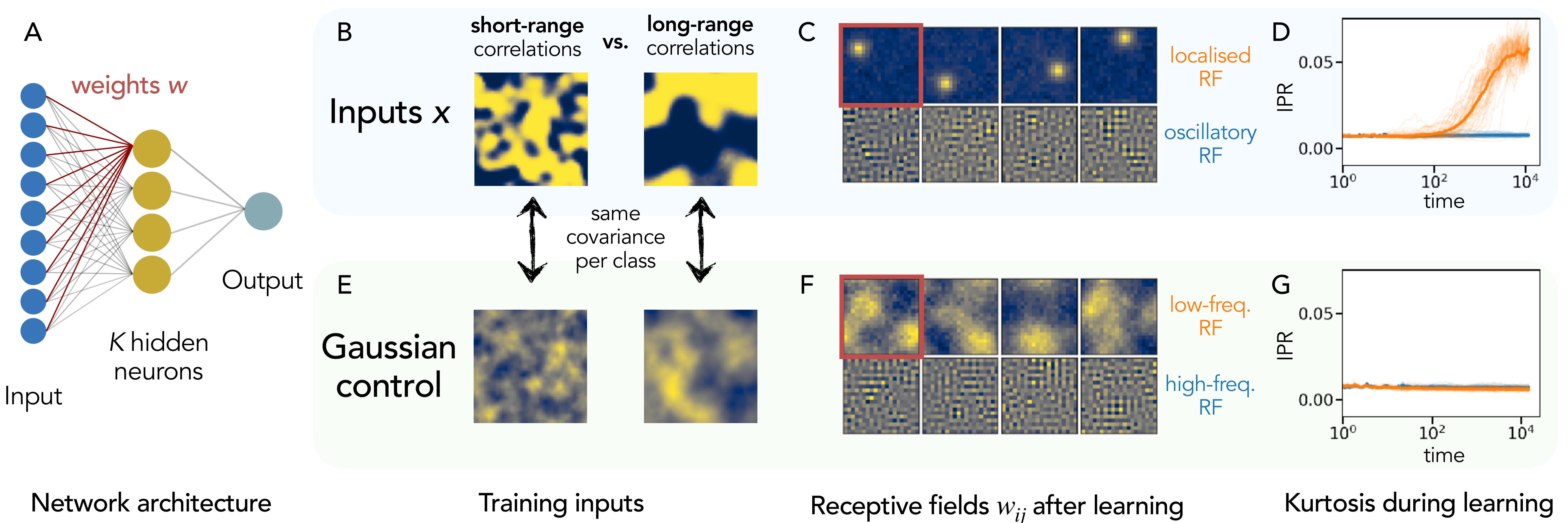}
    \caption{\label{fig:figure1}\textbf{The emergence of convolutional structure in fully-connected neural networks is driven by higher-order input correlations.} \textbf{A} Two-layer, fully-connected neural network with~$K$ neurons in the hidden layer.
      \textbf{B} Networks are trained on a binary classification task with 2-dimensional inputs $\boldsymbol{x}=(x_{ij})$ of size $D=L \times L$ drawn from a translation-invariant random process, \cref{eq:x} with $L=28$. The network has to discriminate inputs with different correlation lengths, $\xi^{-}=0.1L$
      (left) and $\xi^{+}=0.2L$ (right).
      \textbf{C} Receptive fields (RF) of some representative neurons taken from a network with $K=100$ neurons after training. The elements of the each weight vector are arranged in a
      $L\times L$ grid. Half the neurons develop localised receptive
      fields: the magnitude of their weights is significantly different from
      zero only in a small region of the input space. The other neurons converge to superpositions of 2D Fourier components.
      \textbf{D} Inverse Participation Ratio (IPR), \cref{eq:ipr}, of each neuron during training. The IPR is large for localised RF, but remains small for oscillatory RF.
      \textbf{E} Gaussian control dataset: the network is trained on a mixture of two Gaussians, each having zero mean and the same covariance as inputs in~\textbf{B}.
      \textbf{F} Receptive fields after training the network on the Gaussian control data.
      \textbf{G} Inverse participation ratio (IPR),~\cref{eq:ipr}, of the receptive fields of a network trained on Gaussian data.}
  \end{figure*}

In our first experiment, we trained a simple two-layer neural network with $K$ neurons in the hidden layer (\cref{fig:figure1}A) on a synthetic data set with 2-dimensional inputs inputs~$\boldsymbol{x}=(x_{ij})$ of size $D=L \times L$ as in \cref{fig:figure1}B. We generated inputs by first drawing a random vector $\boldsymbol{z}=(z_{ij})$ from a centered Gaussian distribution with a covariance that renders the
input distribution translation invariant along both dimensions.
Each pixel in the synthetic image $x_{ij}$ is then computed as
\begin{equation}
  \label{eq:x}
  x_{ij}=\frac{\psi\left(g z_{ij}\right)}{Z(g)},
\end{equation}
where $\psi(\cdot)$ is a a symmetric, saturating non-linear function such as the error function, $g > 0$ is a gain factor,
and the normalisation constant $Z(g)$ ensures that pixels have unit variance for all values of $g$ (see \methods~for details). Intuitively, the gain factor controls the sharpness in the images: a large gain factor results in images with sharp edges and important non-Gaussian statistics (\cref{fig:figure1}B), while images with a small gain factor are  close to Gaussians in distribution.

Inputs are divided in $M=2$ classes, labelled $y=\pm1$, that differ by the correlation
length~$\xi^\pm$ between pixels: the ``image'' shown on the left of \cref{fig:figure1}B has a shorter
correlation length than the one on the right, hence the input on the left of varies more rapidly in space. The learning task consists in discriminating inputs based on these correlation lengths.

\begin{figure*}[t!]
    \centering
    \includegraphics[width=0.9\textwidth]{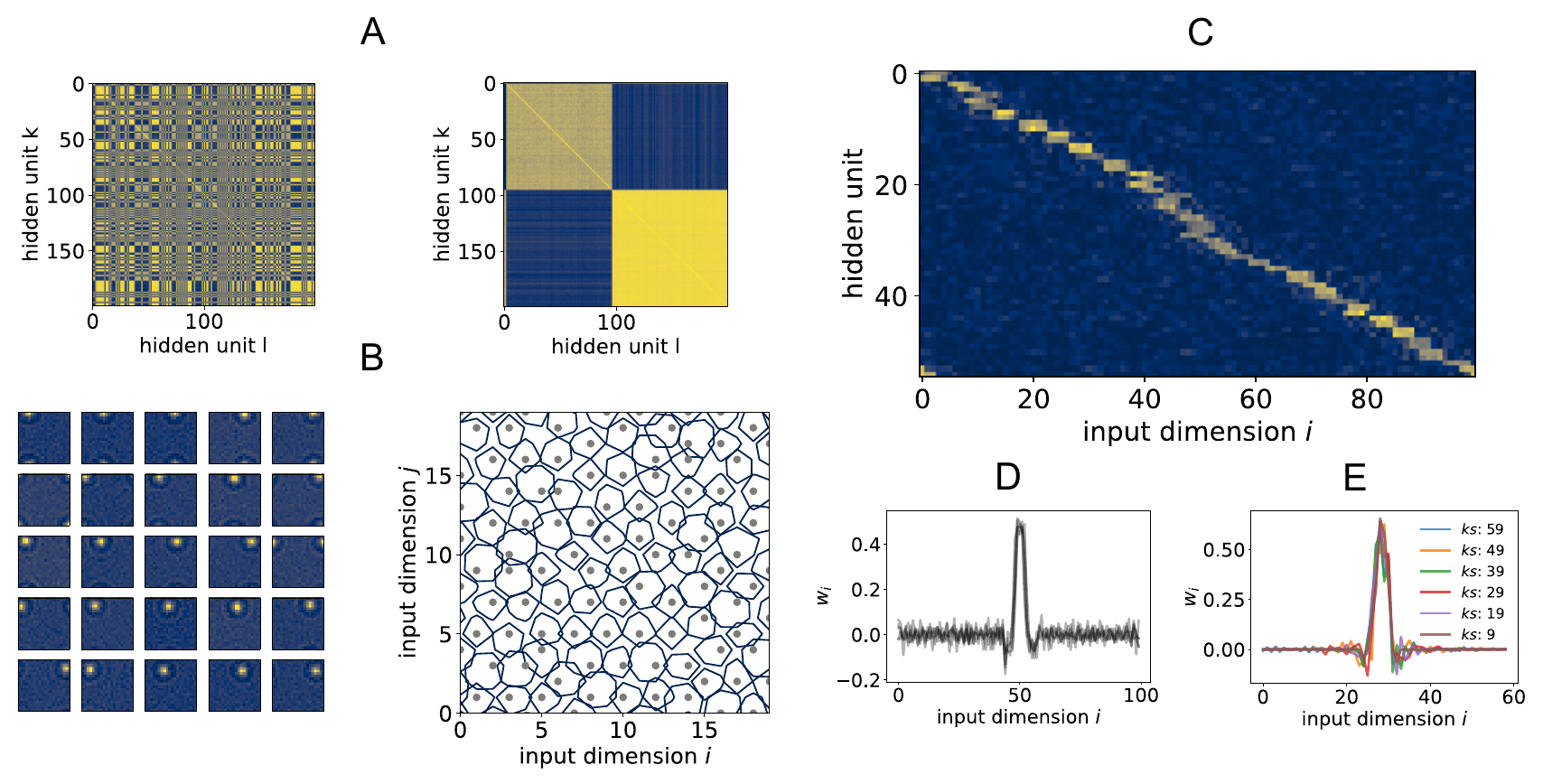}
    \caption{\label{fig:comm_vs_conv}\textbf{Receptive fields of fully-connected networks tile input space and resemble the filters learnt by a convolutional neural network}.
    \textbf{A} Left: Colour plot of the translation-independent distance matrix $\left(d_{kl}\right)$ (see \methods) in a network with $K=200$, trained on a 2-dimensional binary classification task with $\left(\xi^{-},\xi^{+}\right)=\left(\sqrt{2},2\right)$, $L=20$. Right: permuted distance matrix using Hierarchical Clustering, showing how synaptic weight vectors cluster into two groups.
    \textbf{B} Left: Weight intensity of localised receptive fields of a subset of neurons from the network in \textbf{A}. Right: centres (grey) and contour lines (blue) of the whole set of localised RF plotted over the 2-dimensional inputs space.
    \textbf{C} Localised receptive fields in a network with $K=301$ trained on a 1-dimensional task with $\left(\xi^{-},\xi^{+}\right)=\left(\sqrt{10},\sqrt{20}\right)$, $D=L=100$. Weight intensity of each neuron is plotted along the rows, showing that RF are arranged so as to tile the input space. Hidden units were sorted according to their centre, in view of the permutation symmetry.
    \textbf{D} Overlay of five randomly selected receptive fields from \textbf{C}, after centering.
    \textbf{E} Filters of a two-layer convolutional network trained on the same task as \textbf{C}, \textbf{D}. Different colours correspond to different kernel sizes $k_S$, ranging from 9 to 59 pixels.
    \emph{Additional parameters}: gain $g=3$, batch learning with $P=\alpha D$ inputs, $\alpha=10^5$, SGD with batch size $1000$.}
\end{figure*}

A network with $K=100$ hidden neurons reaches $>98\%$ prediction accuracy on this task when trained using online stochastic gradient descent (online SGD), where a new sample $(\boldsymbol{x}, y)$ is drawn from the input distribution at each step of the algorithm. This limit allows us to focus on the impact of the data distribution; we discuss the case of finite training data in \cref{fig:comm_vs_conv}. After learning, the hidden neurons have split into two groups,
with about half the neurons acting as detectors for inputs with long-range correlation.
We plot the weight vector, or the receptive field (RF), of four of these neurons in the top row of
\cref{fig:figure1}C. The receptive fields of these neurons are
\emph{localised}: they only have a few synaptic weights whose magnitude is
significantly larger than zero in a small region of input space.
On the other hand, neurons that detect short-range correlations develop very different representations:
they converge to highly oscillatory patterns, \emph{i.e.}, sparse superpositions of higher frequency Fourier modes.

Beyond the visual inspection of the receptive fields, we can quantify their
localisation by computing the Inverse Participation Ratio (IPR) of their weight
vector $\boldsymbol{w}=(w_i)$,
\begin{equation}
  \label{eq:ipr}
  \mathrm{IPR}(\boldsymbol{w}) = \frac{\sum_{i=1}^D w_i^4}{{\left( \sum_{i=1}^D w_i^2 \right)}^2}.
\end{equation}
The IPR quantifies the amount of non-zero components of a vector. It is commonly used
to distinguish localised from extended eigenstates in quantum mechanics and random matrix theory~\cite{IPRmetz}, and is related to the kurtosis of the weights.
We can successfully employ the IPR to measure the localisation of RF in space throughout learning.
We plot the IPR for the receptive fields of all neurons in \cref{fig:figure1}D as a function of learning time, which is defined as the number of SGD steps divided by the total input size $D$. Localised neurons develop a large IPR over the course of training, while the IPR of neurons with oscillatory receptive fields remains very small.

\subsection*{Higher-order input correlations induce localised receptive fields}
\addcontentsline{toc}{subsection}{Higher-order input correlations induce localised receptive fields}

To determine which of the characteristics of the data set drive the emergence of
localised receptive fields, we trained the same network on a Gaussian
control task (\cref{fig:figure1}E). For each class of inputs, we drew a
new set of control images~$\boldsymbol{c}$ from a Gaussian distribution with the same covariance as the inputs~$\boldsymbol{x}$ from that class. We will sometimes refer to these inputs as the Gaussian process (GP), and denote the non-linear inputs as NLGP. While both the inputs $\boldsymbol{x}$ and the Gaussian controls
$\boldsymbol{c}$ from a given class have the same covariance by construction, and are thus both translation-invariant, the original
inputs $\boldsymbol{x}$ have increasingly sharp edges as we increase the gain factor
$g$. These edges are a visual manifestation of the higher-order spatial
correlations that cannot be captured by the simple Gaussian model. Indeed, the
Gaussian samples appear blurry in comparison to the original data.

The same network with $K=100$ neurons achieved a slightly inferior prediction
accuracy on the Gaussian data set. After learning, the neurons have again split evenly into two populations, detecting short- and long-range correlations, respectively. However, neurons learn very different representations from the data, with example receptive fields shown in \cref{fig:figure1}F. There are no more localised fields; instead, neurons' weights converge to 2D superpositions of low-
and high-frequency Fourier components. This qualitative observation is
borne out by the measurement of the IPR~\eqref{eq:ipr} of the receptive fields, which stays
flat around zero throughout learning, \emph{cf.}~\cref{fig:figure1}G.

Taken together, the results summarised in Fig.~\ref{fig:figure1} show that
localised receptive fields, the first hallmark of convolutions, emerge
autonomously when training two-layer fully-connected networks on a task with
translation-invariant inputs that crucially possess non-Gaussian, higher-order
local structure. This is to be contrasted with other recent studies that focused on the learnability of \emph{tasks} that can be expressed as convolutions in a teacher-student setup~\cite{mei2021learning,misiakiewicz2021learning, favero2021locality}.

\subsection*{Receptive fields tile input space and resemble filters of convolutional networks}
\addcontentsline{toc}{subsection}{Receptive fields tile input space and resemble filters of convolutional networks}

The fully-connected networks we trained also implement weight sharing, the second hallmark of convolutions, where the same filter is applied across the whole input. As shown in \cref{fig:comm_vs_conv}A, hidden units tend to cluster in two distinct groups. These clusters, which are identified by computing similarities between neurons using a translation-invariant measure (see \methods), correspond to neurons with localised and oscillatory receptive fields. These receptive fields were obtained from a network that was trained using SGD on a finite data set with $P=\alpha D$ samples, $\alpha=10^5$.

We show a representative set of neurons with localised RF in \cref{fig:comm_vs_conv}B (left). The centers of
these RF are spread over the input dimensions (\cref{fig:comm_vs_conv}B, right). 
The tiling is more striking in the 1-dimensional case: we show in \cref{fig:comm_vs_conv}C all the localised RF by plotting the weight vectors along the rows of the matrix.
We see that, as the number of hidden neurons $K$ becomes comparable to the input size $D$, the receptive fields tile the input space. A similar tiling has been observed in unsupervised learning with restricted Boltzmann machines by Harsh et al.~\cite{harsh2020placecell}.

We also compared the RF learnt from scratch with the filters learnt in a
two-layer convolutional neural network with different filter sizes trained on the same task~(see \methods~for details).
We found that the learnt convolutional filters are stable across filter sizes (\cref{fig:comm_vs_conv}E).
Strikingly, when a convolutional network is trained on the same task, the
obtained filters strongly resemble the receptive fields learned by
the FC network, as can be seen from a comparison of the filters in
\cref{fig:comm_vs_conv}E with \cref{fig:comm_vs_conv}D, where we show RF of five randomly chosen neurons from \cref{fig:comm_vs_conv}C.

\begin{figure}[t!]
    \centering
    \includegraphics[width=.66\linewidth]{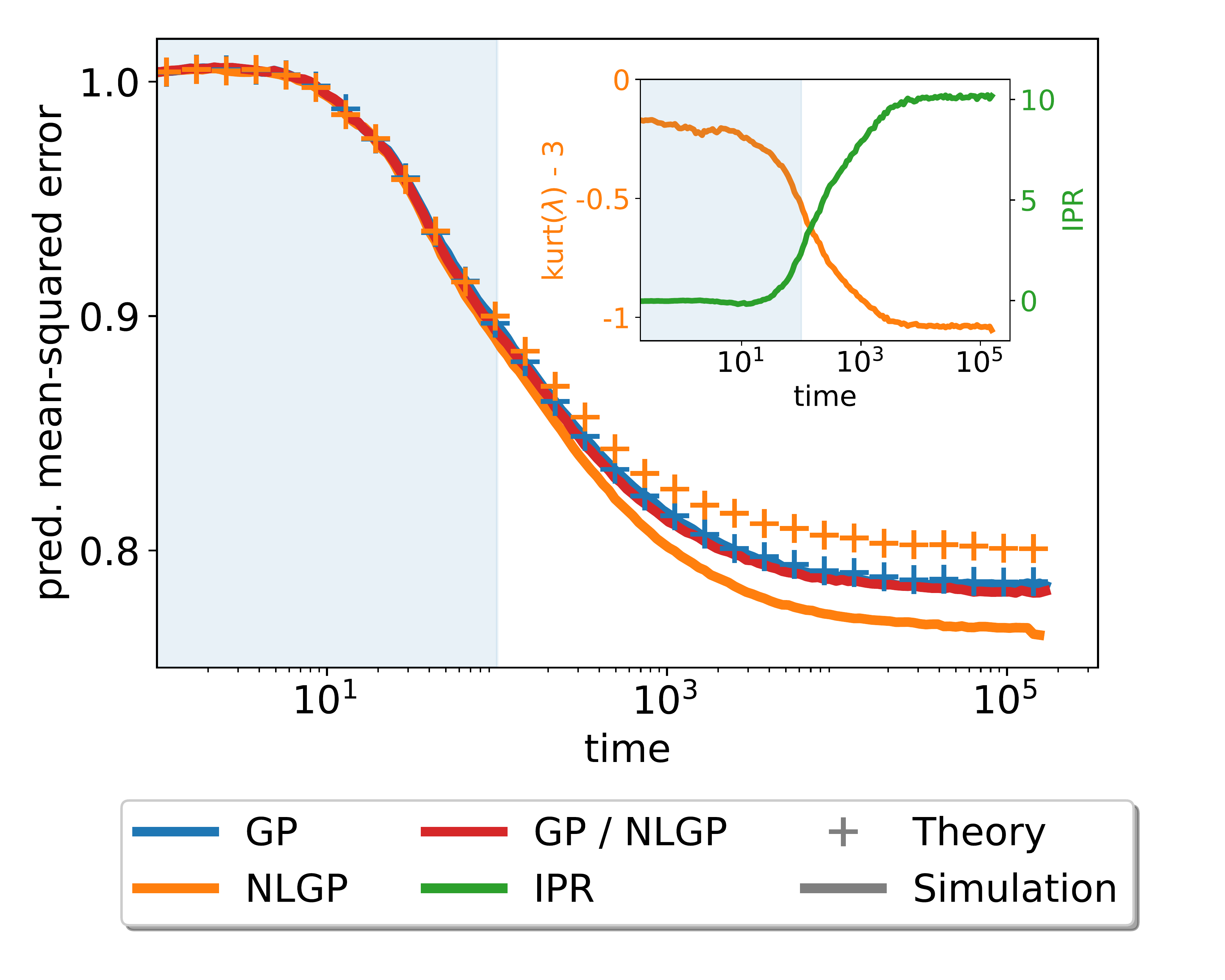}
    \caption{\label{fig:failure-get} \textbf{Existing theories of learning in neural networks break down during the formation of receptive fields.} Prediction mean-squared error $\pmse$~(\ref{eq:pmse}) of a network with $K=8$ neurons trained on non-linear Gaussian inputs (NLGP, \cref{eq:x}, orange) and on the Gaussian control task (GP, blue) with length scales $\xi^+\!=2\xi^-\!=16$. The $\pmse$ is calculated using held-out test data during the simulation (solid lines). We also show the test error of the network trained on GP, but evaluated on NLGP data (GP / NLGP, red). The crosses give the $\pmse$ obtained from evaluating an analytical expression describing the error of an equivalent Gaussian model (\methods). While the analytical expression accurately predicts the error in the beginning of training (blue shaded area), it breaks down for the network trained on NLGP around time~$10^2$. This is precisely the time at which the weights start to localise, as measured by the average IPR~(\ref{eq:ipr}) of the localised weights (inset, green). Simultaneously, the excess kurtosis of the pre-activations of the network decreases (inset, orange). \emph{Additional parameters}: 1-dimensional task with $D=L=400$, learning rate $\eta=0.05$. Curves averaged over twenty runs.}
\end{figure}

\subsection*{Current theories of learning break down during the formation of receptive fields}
\addcontentsline{toc}{subsection}{Current theories of learning break down during the formation of receptive fields}

How can we capture the formation of receptive fields theoretically? There exist precise theories for learning in neural networks with linear activation functions~\cite{baldi1989neural, le1991eigenvalues, krogh1992generalization, saxe2014exact, saxe2019mathematical, advani2020highdimensional}. However, the dynamics of even a deep linear network with several layers will only depend on the input-input and the input-label covariance matrices, i.e.~the first two moments of the data~\cite{saxe2014exact}. This formalism thus cannot capture the formation of receptive fields, which is driven by non-Gaussian fluctuations in the inputs. 
An exact theory describing the learning dynamics is available for non-linear two-layer neural networks with large input size~$D\to\infty$ and a few neurons $K\sim\mathcal{O}(1)$ in the hidden layer~\cite{saad1995a, BiehlLearning}.
We verified that networks in this limit also form receptive fields, see \cref{fig:rf_ode}.
In this limit, one can derive a set of ordinary differential equations that predict the evolution of the (prediction mean-squared) test error $\pmse$ of a network, \eqref{eq:pmse}, when training on Gaussian mixture classification~\cite{refinetti2021classifying}.
In \cref{fig:failure-get}, we show the $\pmse$ of a network with $K=8$ neurons trained on the Gaussian control task (blue lines) and verify that this theory yields matching predictions (blue crosses, full details in \methods).

This type of analysis has recently been extended from mixtures of Gaussians to more complex input distributions thanks to the phenomenon of ``Gaussian equivalence'', whereby the performance of a network trained on non-Gaussian inputs is still well captured by an appropriately chosen Gaussian model for the data. This Gaussian equivalence was used successfully to analyse random features~\cite{liao2018spectrum, seddik2019kernel, mei2021generalization} and neural networks with one or two layers, even when inputs were drawn from pre-trained generative models~\cite{goldt2020modelling, hu2020universality, goldt2021gaussian, loureiro2021capturing}. In \cref{fig:failure-get}, we plot the test error of a network trained on NLGP data together with the theoretical prediction obtained from applying the Gaussian Equivalence Theorem~\cite{goldt2021gaussian} (GET, details are given in \methods). Initially, the theoretical predictions from the GET (orange crosses) agree with the test error measured in the simulation (orange line), but the theory breaks down around time~$\approx 10^2$, when predictions start deviating from simulations.

The breakdown of the Gaussian theory coincides with the localisation of the receptive fields, as measured by their IPR (\cref{eq:ipr}, green line in the inset of \cref{fig:failure-get}). The increased localisation of the weights also coincides with a change in the statistics of the pre-activations of the hidden neurons, $\lambda \sim \sum_i w_i x_i$: the excess kurtosis of $\lambda$ (orange line) is initially close to zero, meaning that $\lambda$ is approximately Gaussian, but decreases as the weights localise, indicating a transition to a non-Gaussian distribution.

We can finally see from \cref{fig:failure-get} that the network is only influenced by the second-order fluctuations in both the NLGP and the GP at the beginning of training, since the $\pmse$ for models trained on NLGP and GP initially coincide. Likewise, a network trained on GP and evaluated on NLGP test data has the same test accuracy as the network trained directly on NLGP in the early stages of learning (red line). The higher-order moments of the NLGP inputs start influencing learning only at a later stage, when the IPR of the weight vectors increases and the Gaussian theory breaks down. This sequential learning of increasingly higher-order statistics of the inputs is reminiscent of how neural networks learn increasingly complex functions during training. Simplicity biases of this kind have been analysed in simple models of neural networks~\cite{schwarze1992generalization, saad1995a, engel2001statistical, SaxeGanguliMathematical, rahaman2019spectral} and have been demonstrated in modern convolutional networks~\cite{FunctionIncreasingComplexity}. The sequential learning of increasingly higher-order statistics and the ensuing breakdown of the GET to describe learning is a result of independent interest which we will investigate further in future work.

The failure of the Gaussian theory to describe the emergence of receptive
fields forces us to develop a new theoretical approach. We make
a first step in this direction by introducing a simplified model,
which allows us to analyse the impact of the non-Gaussian statistics.

\subsection*{A simplified model highlights the importance of non-Gaussian statistics}
\addcontentsline{toc}{subsection}{A simplified model highlights the importance of non-Gaussian statistics}

\begin{figure*}[ht]
    \centering
    \includegraphics[width=\textwidth]{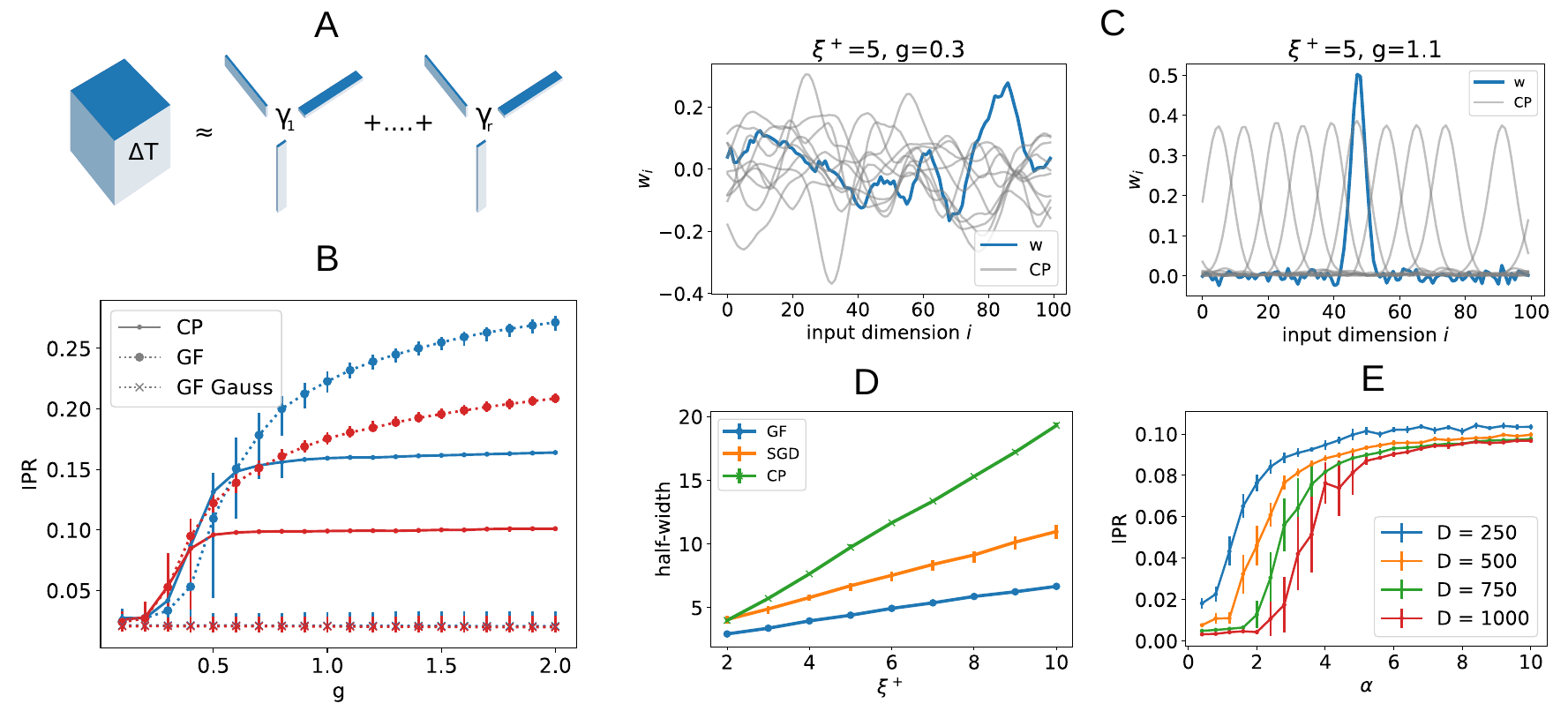}
    \caption{\label{fig:flow_and_cp}
    \textbf{Non-gaussianity drives pattern-formation in a simplified model of gradient descent dynamics.}
    \textbf{A} Pictorial illustration of CP decomposition~\cite{kiers1998three, KoldaTensor}, a tensor decomposition technique where a tensor (here a three-way tensor) is decomposed into a weighted sum of rank-1 tensors.
    \textbf{B} Inverse Participation Ratio (IPR) as the gain factor $g$ of the data is increased, thereby increasing the non-Gaussianity  of the inputs (cf.~\cref{eq:x}). The IPR is shown for i) the leading CP factor of the fourth-order cumulant $\Delta T^2$ (solid), ii) the weight vector of a single neuron obtained by integrating the gradient flow (GF) equation (\cref{eq:reduced_model_cumulant-main}, dots) and iii) the weight vector obtained by integrating the GF equation where higher-order moments have been replaced with Gaussian moments of the inputs (crosses). Error bars indicate interquartile range around median across $30$ samples.
    \textbf{C} Synaptic weight vectors $\boldsymbol{w}$ (blue) obtained from integrating the GF equation for small and large values of the gain parameter (left and right, resp.). In grey, we show the ten leading CP factors $\boldsymbol{u}_k$ of the fourth-order cumulant $\Delta T^2$ for both datasets.
    \textbf{D} Half-width of the weight vector obtained from integrating the GF equation~\cref{eq:reduced_model_cumulant-main} (blue) and of the leading CP factor of the fourth-order cumulant (green), as the correlation length $\xi^{+}$ of the inputs is increased. We also show the half-width of the weight vector obtained by training the simplified model directly using SGD updates with finite step size (orange).
    \textbf{E} Maximal IPR among the first $20$ CP factors for a dataset containing $P=\alpha D$ inputs with $g=3$, $\xi=5$, for increasing $\alpha$ and size $D$. Error bars indicate $\pm 3$ std error around the average across $30$ samples.
    \emph{Additional parameters in \textbf{A}-\textbf{D}}: 1-dimensional inputs, $D=L=100$, $K=1$, $\xi^{-}=0$, cumulants estimated from a dataset with $P=\alpha D$ inputs, $\alpha=100$, learning rate $\eta=0.01$, bias fixed at $b=-1$. }
\end{figure*}

The analysis of a reduced model with a single neuron reveals an interesting
connection between the higher-order statistics of the data and the
pattern-formation mechanism driving the emergence of convolutional structure. We consider a single neuron with a polynomial activation function $\tilde{\sigma}$ of order~3 and study the weight updates of stochastic gradient descent in the limit of small learning rate. This leads us to consider the gradient flow (GF) dynamics of the neuron's weight vector $\boldsymbol{w}$ after averaging over the data distribution, that takes the form
\begin{equation}
  \label{eq:reduced_model-main}
  \dot{\boldsymbol{w}} = \frac{1}{M} \sum_{\mu=1}^{M} \left[c_{2}^{\mu}\left(y^{\mu},v,b\right)C^{\mu}\boldsymbol{w}+c_{4}\left(v,b\right)T^{\mu}\boldsymbol{w}^{\otimes 3}\right],
\end{equation}
where $C_{ij}^{\mu}=\left\langle x_{i}^{\mu}x_{j}^{\mu}\right\rangle $ and
$T_{ijk\ell}^{\mu}=\left\langle
  x_{i}^{\mu}x_{j}^{\mu}x_{k}^{\mu}x_{\ell}^{\mu}\right\rangle $ are the second-
and fourth-order joint moments of the inputs in the $\mu$th class, respectively
($M=2$). The notation $\boldsymbol{w}^{\otimes 3}$ indicates a three-fold outer product of the vector $\boldsymbol{w}$ with itself, see \cref{eq:ttsv-methods}. In \cref{eq:reduced_model-main}, we discarded the 5th order term and introduced the coefficients $c_2$, $c_4$, as described in \methods.
The same steps can be used to derive a similar GF equation for the bias of the neuron.
Since the results do not depend on the exact value of the bias or on its dynamics, here we simplify the discussion by fixing the bias at $b=-1$. We verified that following the gradient flow dynamics of the model~\eqref{eq:reduced_model-main} yields a localised receptive field (cf.~the blue lines in \cref{fig:flow_and_cp}).

While a complete analysis of the synaptic dynamics for generic fourth-order
tensors $T^{\mu}$ is very complicated and beyond our scope, we can gain insight
by rewriting $T^{\mu}$ as a sum of the cumulant $\Delta T^{\mu}$ plus the contribution from the second order moment
\begin{equation}
  \label{eq:cumulant}
  T^{\mu}_{ijk\ell} = \Delta  T^{\mu}_{ijk\ell} + C^{\mu}_{ij} C^{\mu}_{k\ell} + C^{\mu}_{ik}C^{\mu}_{j\ell} + C^{\mu}_{i\ell} C^{\mu}_{jk}.
\end{equation}
The cumulant $\Delta T^\mu$ has the useful property that it is exactly zero for
Gaussian inputs; in other words, it quantifies the non-Gaussian part of the
fourth-order input statistics. We can then rewrite the synaptic dynamics as
\begin{equation}
\label{eq:reduced_model_cumulant-main}
\dot{\boldsymbol{w}} =\frac{1}{M}\sum_{\mu=1}^M\left(c_{2}^{\mu}+c_{4}q^{\mu}\right)C^{\mu}\boldsymbol{w}+\frac{c_{4}}{M}\sum_{\mu=1}^M\Delta T^{\mu}\boldsymbol{w}^{\otimes 3},
\end{equation}
where $q^{\mu}= \boldsymbol{w}^{T}C^{\mu}\boldsymbol{w}$ is the so-called self-overlap of the synaptic weight, and we dropped the dependence in $c_2$ and $c_4$ for brevity. 

In our data model, the relative importance of the non-Gaussian
statistics in the inputs is controlled by the gain factor $g$ introduced in
\cref{eq:x}: for small values of $g$, the error function is almost linear, and the inputs almost Gaussian. For Gaussian inputs, the cumulant $\Delta T^\mu=0$ and the synaptic dynamics~$\dot{\boldsymbol{w}}$ are thus given only by the first term on the right-hand side. It can be shown that the fixed point equations imply a very sparse power-spectrum (\methods); in other words, the weights converge to a superposition of only a few Fourier components, in line with our general finding for $K\geq 1$.

We train the single neuron on a task with non-linear inputs (NLGP) at various values of the gain factor $g$. We set the correlation length of inputs in one class to zero, $\xi^{-}=0$, and vary the correlation length for the second class.
Integrating \cref{eq:reduced_model_cumulant-main} yields the weight vector of
the neuron at the end of training. We plot the IPR of this weight as a function
of the gain factor $g$ with the dotted line in \cref{fig:flow_and_cp}B (blue and
red for $\xi^+=3,\;\xi^+=5$, respectively), and show the weight for two values of the gain factor
(\cref{fig:flow_and_cp}C).
The single neuron develops a receptive field that is increasingly localised as the gain factor, and hence the non-Gaussianity of the inputs, increases. We also integrated \cref{eq:reduced_model_cumulant-main} keeping only the first term, so as to only retain the influence of the Gaussian part of the data on the learning dynamics. Integrating the reduced equation yields a synaptic weight that is not localised - its IPR is negligible for all values of the gain
(crosses in \cref{fig:flow_and_cp}B). The driving force behind
the emergence of localised receptive fields in the reduced model is thus the
fourth-order cumulant $\Delta T^\mu$.

We can gain insight into the structure of these higher-order correlations by
means of tensor decomposition. Just like a matrix (which is a tensor of order 2) can be decomposed into a sum of outer products between eigenvectors,
higher-order tensors can be expressed as a sum of a relatively small number of
outer products of vectors, which are called \emph{factors} in this context.
There exist several ways to decompose a tensor; here we focus on the PAFAC/CANDECOMP (CP) decomposition of the fourth-order cumulant, whereby
\begin{equation}
    \Delta T = \sum_{k=1}^{r} \gamma_k \boldsymbol{u}_k \otimes \boldsymbol{u}_k \otimes \boldsymbol{u}_k \otimes \boldsymbol{u}_k
\end{equation}
and $r$ is the \emph{rank} of the decomposition
(see \cref{fig:flow_and_cp}A for an illustration of a third-order tensor).
Tensor decomposition has been successfully applied to supervised and unsupervised machine learning~\cite{AnandkumarTensor, KoldaTensor}, and its relevance in the context of unsupervised synaptic plasticity has also been recently recognised: Ocker and Buice~\cite{ocker2021tensor} showed that a polynomial version of Hebbian learning can recover the dominant tensor eigenvectors of higher-order correlations, in a way that is similar to how the classic Oja rule recovers the leading eigenvector of the input covariance~\cite{oja1982simplified}.

We find that the progressive localisation of the receptive fields mirrors the
localisation of the dominant CP factors of the fourth-order cumulant (full curves in \cref{fig:flow_and_cp}B and C). 
For large enough data sets and high rank $r$, the CP factors of the inputs tile the input space in a manner similar to the receptive fields obtained on the supervised task with a large number of hidden neurons $K$ (\cref{fig:flow_and_cp}C, right). The precise shape of the receptive fields is controlled by the CP factor obtained from a rank~$r=1$ decomposition of the fourth-order cumulant: as we show in \cref{fig:flow_and_cp}D, the half-width of the localised receptive field obtained using both the reduced model (gradient flow) and the full SGD dynamics closely follow the half-width of the CP factor when the correlation length~$\xi^{+}$ is varied. Since the CP factor is computed over the correlated inputs, while the perceptron sees both correlated and uncorrelated inputs, the half-width of the receptive fields obtained from learning are smaller than that of the CP factor. Furthermore, the half-width of the receptive field obtained from SGD is closer to the value of the CP factor, as the gradient flow dynamics only has access to the first four moments of the inputs, cf.~\eqref{eq:reduced_model_cumulant-main}, while the perceptron trained with SGD sees all the moments of its inputs.

It is also attractive to relate pattern formation in weight space with bump
attractor dynamics in models with non-linear local interactions
\cite{AmariDynamics, RedishAttractor, CompteSynaptic}. The dynamics of
\cref{eq:reduced_model_cumulant-main} in the presence of a low-rank CP
decomposition of $\Delta T^\mu$ is instructive, in that it manifests
attractor-like phenomenology (Fig.~S2). This kind of dynamics in weight space is reminiscent of memory retrieval in continuous Hopfield models, where a third
order interaction among spin variables mediated by the fourth-order moment
tensor is necessary for retrieval \cite{BolleSpherical}. When the previously
introduced polynomial activation function $\tilde{\sigma}$ is used in
conjunction with finite batch training and a plastic readout weight, both drifting periods and transitions between localised fields are apparent over the course of learning, as a result of effective noise induced by finite batch size (Fig. S3). Although abrupt transitions are accompanied by transient sweeps of the readout weight $v$, both $v$ and the bias $b$ remain approximately constant while the position or sign of the localised field change, as predicted by the symmetry in the training data. Drifting localised fields has also been observed when a generative model (Restricted Boltzmann Machine) is trained using Contrastive Divergence \cite{harsh2020placecell} to reproduce configurations from a 1-dimensional Ising chain.

\section*{Discussion}
\addcontentsline{toc}{section}{Discussion}

Convolutional neural networks achieve better performance and need less samples than fully-connected (FC) networks when trained with stochastic gradient descent, especially in vision, even though sufficiently wide FC networks can express convolutions. However, FC networks do not develop a convolutional structure autonomously when trained on a supervised image classification task. d'Ascoli et al.~\cite{dascoliNeedle} recently highlighted the dynamical nature of this problem when they showed that convolutional
solutions are not reachable by SGD starting from random fully-connected initial conditions. The training has to be augmented by techniques like weight pruning~\cite{pellegrini2021sifting} or complex regularisation schemes~\cite{NeyshaburScratch} in order to learn weight matrices that display local connectivity and are organised in patterns reminiscent of convolutional networks.

Here, we showed that convolutional structure in fully connected neural networks can emerge during training on a supervised learning task. We designed a minimal model of the visual scene whose non-Gaussian, higher-order statistics are the crucial ingredient for the emergence of receptive fields characterised by both localisation and weight sharing.
We further highlighted the dynamical nature of the learning phenomenon: the progression from second-order to higher-order statistics during learning is an example of how neural networks learn functions of increasing complexity. 

We studied the pattern-formation mechanism of localised receptive fields using a reduced model with a single neuron.
A similar approach has recently been used in the context of unsupervised learning of configurations generated by lattice models in physics~\cite{harsh2020placecell}, where weight localisation was interpreted in terms of a Turing instability mechanism. Our work follows the legacy of earlier pioneering studies on single neurons that analysed storage and memory retrieval of spatially correlated~\cite{Monasson_correlatedpatterns, Monasson_properties, TarkowskiCorrelatedMemories, TarkowskiPerceptron} and invariant datasets~\cite{TarkowskiInvariant}. At variance with these classical works, here we focused on the dynamics of learning and studied the role of higher-order statistics.
Encapsulating these higher-order information in appropriate order-parameters presents itself as a crucial next step, in that it will allow studying the typical structure of the optimal solution to supervised learning problems with complex datasets.

The analysis of the single-neuron model led us to relate the emergence of structural properties of network connectivity to the tensor decomposition of higher-order input cumulants. For a neural network to be able to capture this structure, a large amount of data must be processed: one indeed expects the structure in the dominant CP factors (or tensor eigenvectors) of higher-order moments to depend on the number of samples. We demonstrate this point numerically in \cref{fig:flow_and_cp}E, where we plot the localisation of the dominant CP factors of the fourth-order cumulant (measured by their IPR) as a function of the number of samples for various input sizes $D$ -- details on how we performed tensor decomposition for large $D$ are given in Sec.~D of the Supporting Information. We note that as $D$ increases (for constant correlation length $\xi$) the sample fluctuations increase at the transition. A better understanding of this transition and similar other properties of higher-order cumulants represents an interesting direction for further study. We expect that typical-case studies of the decomposition of large random tensors~\cite{richard2014statistical, montanari2015limitation,lesieur2017statistical,chen2019phase, perry2020statistical} and approaches based on random matrix theory~\cite{goulart2021random} will prove fruitful in this direction, similar to the progress in understanding the spectral properties of random covariance matrices~\cite{baik2005phase, auffinger2013random, potters2020first}. 

The extension of our single-neuron model to the multi-neuron case proves complicated due to the effective repulsive interactions between different weight vectors that appear even for Gaussian inputs. However, studying these effective repulsive interactions in the general case is an interesting future direction for our work, as it could shed light on the mechanism of space-tiling.

While in this work we focused on translation symmetry, recent developments in deep learning have dealt with a variety of symmetry groups~\cite{CohenGroupEquivariant, KondorCompact, SteerableCNN} and a general framework has been introduced for constructing equivariant layers capable of dealing with input invariances~\cite{Bronstein2021geometric}. It is thus natural to consider the impact of a generic symmetry group on the higher-order statistics of the data and ask for the conditions under which such a structure is learnable, and whether there exists a minimal amount of data necessary to detect such an invariance.

In the interest of tractability, here we employed a synthetic data model and a simple two-layer network. A full dynamic analysis of deeper architectures is complicated by the highly nonlinear dynamics of gradient descent, already evident even in linear networks~\cite{saxe2014exact, saxe2019mathematical, advani2020highdimensional}. Understanding how invariances in data interact with depth in a neural network is an interesting direction of future investigation, both at the analytical and numerical level.

Studying the formation of receptive fields in recurrent networks is an interesting future direction for two reasons: recurrent networks provide an effective tool for capturing the spatio-temporal dynamics of the visual scene~\cite{LiangRecurrent, NicoRecurrentBetter}, and they are promising models for the processing stages in the visual system~\cite{NicolRecurrentFlexible, NicoCircles, DiCarloFastRecurrent}. The simplest test-bed for our approach would be the study of recurrent networks solving classification tasks~\cite{FarrellRecurrent} in the presence of data invariances.

Finally, we note that our work establishes an intriguing connection between supervised and unsupervised learning. We found that the gradient updates drive the weights in directions that increase the non-Gaussianity of the pre-activations of the hidden neurons, as measured by their excess kurtosis (cf.~\cref{fig:failure-get}). The representations found in this way perform better than the ones obtained from Gaussian inputs. Maximising non-Gaussianity has long been recognised as a powerful mechanism to extract meaningful representations from images, e.g.~kurtosis maximization in independent component analysis~\cite{bell1996edges, hyvarinen2000independent}. This work represents a first step towards linking generative model approaches to vision with task-relevant feature extraction carried out by supervised learning rules.

\section*{Methods}
\addcontentsline{toc}{section}{Methods}

\phantomsection
\subsection*{Data models}
\addcontentsline{toc}{subsection}{Data models}

Our data set consists of inputs $\boldsymbol{x}$ that can be one- or two-dimensional, divided in $M$ distinct classes. Here, we illustrate the different types of inputs in one dimension. 

A data vector of the \textbf{non-linear Gaussian process (NLGP)} is given by
$\boldsymbol{x}^{\mu}=Z^{-1}(g)\psi\left(g\boldsymbol{z}^\mu\right)$, where
$\boldsymbol{z}^\mu$ is a zero-mean Gaussian vector of length $L$ and covariance
matrix
\begin{equation}
  C_{ij}^\mu = \left\langle z_{i}^{\mu}z_{j}^{\mu}\right\rangle
  =e^{-\left(\left|i-j\right|\right/\xi^\mu)^2},
\end{equation}
with $i,j = 1,2,\ldots,L$. The
covariance thus only depends on the distance between sites $i$ and $j$, given by~$\left|i-j\right|$. The normalisation factor $Z(g)$ is chosen such that
$\Var\left(x\right)=1$. Throughout this work, we took~$\psi$ to be a symmetric
saturating function
$\psi\left(z\right)=\erf\left(\nicefrac{z}{\sqrt{2}}\right)$, for which
$Z(g)^2=\nicefrac{2}{\pi} \arcsin\left(g^2 / (1 + g^2)\right)$. We also enforce
periodic boundary conditions.

We create the \textbf{Gaussian clone (GP)} by drawing inputs from a Gaussian distribution with mean zero and the same covariance as the corresponding NLGP. The covariance of the NLGP can be evaluated analytically for $\psi\left(z\right)=\erf\left(\nicefrac{z}{\sqrt{2}}\right)$ and reads
\begin{equation}
   \label{eq:gp-covariance}
    \langle x_i^\mu x_j^\mu \rangle = \frac{2}{\pi Z(g)} \mathrm{arcsin}\left(\frac{g^2}{1+g^2}C^\mu_{ij}\right)
\end{equation}
where we have used that fact that $C_{ii}=1$. The experiments on Gaussian processes (GP) are thus \emph{not} performed on the Gaussian variables $\boldsymbol{z}$; they are performed on Gaussian random variables with covariance given in \cref{eq:gp-covariance}. In this way, we exclude the possibility that the change in the two-point correlation function from applying the non-linearity $\psi$ is responsible for the emergence of receptive fields.

For 1-dimensional inputs, the fact that the covariances of the NLGP and the GP  
depend only on the distances between pixels $|i-j|$ implies that they are \emph{circulant matrices}~\cite{horn2012matrix}.
These matrices display a number of useful properties: they can be diagonalised using discrete Fourier
Transform (DFT), and thus any two circulant matrices of the same size
can be jointly diagonalised and commute with each other. We use this fact
in the analysis of the reduced model to diagonalise the dynamics of the synaptic
weights (see ``Gaussian inputs'' below).

We obtain the covariance for 2-dimensional inputs by taking the Kronecker product of the one-dimensional covariance matrix with itself. For any dimension, we indicate the total input size by~$D$.

\subsection*{Details on neural network training}
\addcontentsline{toc}{subsection}{Details on neural network training}

We trained a two-layer fully connected (FC) network with $K$ hidden unit and activation function $\sigma$. The output of the network to an input $\boldsymbol{x}$ is:
\begin{equation}
  \label{eq:2lnn}
    \phi\left(\boldsymbol{x}\right)=\sum_{k=1}^{K}v_{k}\sigma\left(\sum_{i=1}^D w_{ki} x_{i}+b_{k}\right),
\end{equation}
with $W\in \mathbb{R}^{K\times D}$ the matrix of first layer weights and $b_k$ the hidden unit biases. We initialised $W$ with i.i.d. zero-mean Gaussian entries with variance $\nicefrac{1}{D}$. To obtain a minimal model of developing convolutions, we fixed the second layer weights of the network to the value $v_k=\nicefrac{1}{K}$. We show that the emergence of receptive fields also occurs in networks where both layers are trained from scratch in the Supporting Information (Fig. S1). We employed the sigmoidal activation function $\sigma(h)=\erf\left(\nicefrac{h}{\sqrt 2}\right)$ for the results shown in the main text, and verified that localised receptive fields also emerge with the ReLU activation~$\sigma(x) = \max(0, x)$.

We trained the network using vanilla stochastic gradient descent (SGD),
using both standard mini-batch learning from a finite dataset and online learning. In the latter, a new sample $(\boldsymbol{x}, y)$ is drawn from the input distribution for each step of SGD. This limit is widely used in the theory of neural networks, as it permits focusing on the statistical properties of the inputs, without effects that could arise due to scarce amounts of data. It has furthermore been shown that online learning is quite close to the
practice of deep learning, where heavy data augmentation schemes lead to very large effective dataset sizes~\cite{nakkiran2020bootstrap}.

For binary discrimination tasks, we used $\left\{-1,+1\right\}$ output for the two classes. We focus our analysis on mean-square loss for simplicity of mathematical treatment. We verified that cross-entropy loss does not alter our main results, cf.~\cref{sec:relu-cross-entropy}.
For the comparison with convolutional networks, we employed a two-layer network composed of a convolutional layer with circular padding, followed by a fully connected layer with linear output.

\subsection*{Invariant overlap and clustering}
\addcontentsline{toc}{subsection}{Invariant overlap and clustering}

In order to compare different weight vectors $\boldsymbol{w}_k$ (rows of the first-layer weight matrix $W$), we introduce a similarity measure that is invariant to translation. Given two normalized weight vectors $\boldsymbol{w}_k$  and $\boldsymbol{w}_l$, the overlap $\tilde{q}_{kl}$ reads:
\begin{equation}
\tilde{q}_{kl}=\frac{1}{D}\left|\tilde{\boldsymbol{w}}_{k}\right|\cdot\left|\tilde{\boldsymbol{w}}_{\ell}\right|,
\end{equation}
where $\tilde{w}_{k\tau}$ stands for the $\tau$th Fourier components of the vector $\boldsymbol{w}_{k}$ and the absolute value is computed entrywise. The latter operation makes $\tilde{q}_{kl}$ invariant with respect to translation by removing the phase information.
To help identify the set of localised receptive fields, we cluster the weight vectors $\boldsymbol{w}_k$ using a distance matrix $d_{kl}=1 - \tilde{q}_{kl}$ and (average-linkage) Hierarchical Clustering. The same procedure is employed both in 1 and 2 dimensions.

\subsection*{Gaussian equivalence}
\addcontentsline{toc}{subsection}{Gaussian equivalence}

The prediction mean-squared test error for a given network (as shown in \cref{fig:failure-get}) is defined as
\begin{equation}
  \label{eq:pmse}
  \pmse \equiv \langle {\left( \phi(\boldsymbol{x}) - y \right)}^2 \rangle_{\boldsymbol{x},y}.
\end{equation}
The average is taken over the data distributions $(\boldsymbol{x}, y)$. We compute this error during the simulation by evaluating the performance of the model on held-out test data. A crucial observation is that the inputs~$\boldsymbol{x}^\mu$ only affect the network output~\eqref{eq:2lnn} via the dot-product with the network's weights pre-activations
\begin{equation}
    \label{eq:lambda}
    \lambda_{k}^{\mu}=\sum_{i=1}^D w_{ki} x_{i}^{\mu}.
\end{equation}
The high-dimensional average over inputs in \cref{eq:pmse} can thus be replaced by a low-dimensional average over the pre-activations. This approach to studying the learning dynamics of two-layer networks was pioneered by Saad \& Solla~\cite{SaadSollaExact} and Riegler \& Biehl~\cite{RieglerOnline}, who studied neural networks learning random functions of i.i.d.~Gaussian inputs, and was recently made rigorous~\cite{goldt2019dynamics, veiga2022phase}. To obtain the theoretical predictions in \cref{fig:failure-get}, we built on a recent extension of this analysis to the case of mixtures of Gaussian inputs with non-trivial input correlations~\cite{refinetti2021classifying}.

For Gaussian inputs (GP), the $K$ pre-activations $\lambda_k^{\mu}$ are jointly Gaussian for each class $\mu$. For non-Gaussian inputs (NLGP), one can invoke the Gaussian Equivalence Theorem (GET), which stipulates that, for a wide class of input distributions, the pre-activations $\lambda_k^{\mu}$ remain Gaussian~\cite{goldt2021gaussian, hu2020universality}. In \cref{fig:failure-get}, we evaluate the test error of a network trained on NLGP by replacing the actual pre-activations $\lambda_k^\mu$ with Gaussian random variables $\tilde{\lambda}_k^{\mu}$ of mean zero and covariance $\langle \tilde{\lambda}_k^{\mu} \tilde{\lambda}_\ell^{\mu} \rangle = w_k C^{\mu} w_\ell$ for each input class $\mu$. The GET prediction for the test error is then obtained by evaluating the average in \cref{eq:pmse} over the Gaussian variables $\tilde{\lambda}^{\mu}_k$. As we show in \cref{fig:failure-get}, the predictions based on Gaussian equivalence match the simulation initially, but break down when the IPR of the localised weights increases. 

The influence of the localisation of the weight~$w_k$ on the higher-order statistics of the local fields $\lambda_k^\mu$ can be seen by computing the \emph{statistical} excess kurtosis
\begin{equation}
    \label{eq:kurtosis}
    \mathrm{kurtosis}(\lambda^\mu_k) = \frac{\langle \lambda_k^4\rangle_\mu}{{\langle \lambda_k^2\rangle_\mu}^2} - 3,
\end{equation}
where the average is taken over the $\mu$th input class. In the inset of \cref{fig:failure-get}, we plot the excess quantity averaged over the neurons with localised weights and over the two input classes.

\subsection*{Reduced model and tensor decomposition}
\addcontentsline{toc}{subsection}{Reduced model and tensor decomposition}

In order to analyze the learning dynamics in the presence of higher-order
statistics, we expand the activation function to third order
$\sigma\left(h\right)\approx\tilde{\sigma}\left(h\right)=\alpha_1
h-\frac{\alpha_3}{3} \; h^3$.
In particular, for $\sigma\left(h\right)=\erf\left(\nicefrac{h}{\sqrt{2}}\right)$ one has $\alpha_1=\sqrt{\nicefrac{2}{\pi}}$ and $\alpha_3=\nicefrac{1}{\sqrt{2\pi}}$. Upon presentation of a pattern from the $\mu$th class, the gradient update reads $\Delta w_{i}^{\mu}=\eta v\left(y^{\mu}-vg\left(h^{\mu}\right)\right)g^{'}\left(h^{\mu}\right)x_{i}^{\mu}$, with~$\eta$ a small learning rate.
We thus get:
\begin{equation}
\Delta w_{i}^{\mu}\propto c_{2}^{\mu}\left(y^{\mu},v,b\right)\sum_{a=1}^D x_{a}^{\mu}w_{a}+c_{4}\left(v,b\right)\sum_{abc}^D x_{a}^{\mu}x_{b}^{\mu}x_{c}^{\mu}x_{i}^{\mu}w_{a}w_{b}w_{c},
\end{equation}
where we discarded the 5th order term in $\boldsymbol{w}$ and set:
\begin{align*}
 & c_{2}^{\mu}\left(y^{\mu},v,b\right)=-v\left[\left(\alpha_1-\alpha_3 b^{2}\right)^{2}+2\alpha_3 b\left(y^{\mu}-\alpha_1 vb+\frac{\alpha_3}{3}vb^{3}\right)\right]\\
 & c_{4}\left(v,b\right)=v\left[\frac{4}{3}\alpha_3\left(\alpha_1-\alpha_3 b^{2}\right)-2\alpha_3^{2}b^{2}\right].
\end{align*}
Averaging over the inputs $\boldsymbol{x}^{\mu}$ and summing across the $M$ classes, we get
\begin{equation}
\left\langle \Delta \boldsymbol{w}\right\rangle =\frac{\eta}{M}\sum_{\mu=1}^{M}\left(c_{2}^{\mu}C^{\mu}\boldsymbol{w}+c_{4}T^{\mu}\boldsymbol{w}^{\otimes 3}\right),
\label{eq:reduced_model-methods}
\end{equation}
where we dropped the dependence in $c_2$ and $c_4$ for brevity. Recall that $C_{ij}^{\mu}=\left\langle x_{i}^{\mu}x_{j}^{\mu}\right\rangle$ and $T_{ijk\ell}^{\mu}=\left\langle x_{i}^{\mu}x_{j}^{\mu}x_{k}^{\mu}x_{\ell}^{\mu}\right\rangle $ are the second- and fourth-order joint moments of the inputs in the $\mu$th class, respectively. Here and in the main text we used the notation $T\boldsymbol{w}^{\otimes 3}$ to indicate the 3-times product of the tensor $T$ with the vector $\boldsymbol{w}$:
\begin{equation}
\left(T\boldsymbol{w}^{\otimes 3}\right)_i = \sum_{abc}^D T_{abci} w_a w_b w_c.
\label{eq:ttsv-methods}
\end{equation}

\subsubsection*{Gaussian inputs}
\addcontentsline{toc}{subsubsection}{Gaussian inputs}

Using Wick's theorem for centered data, $T_{abcd}=C_{ab}C_{cd}+C_{ac}C_{bd}+C_{ad}C_{bc}$, we can express the third-order term using the respective covariance matrices:
\begin{equation}
\sum_{abc}^D T^{\mu}_{abci}w_{a}w_{b}w_{c}=3q^{\mu}\sum_{c=1}^D C^{\mu}_{ci}w_{c},
\end{equation}
with $q^{\mu}=\boldsymbol{w}^T C^{\mu}\boldsymbol{w}$ the single-unit definition of the overlap. In the limit of small learning rate $\eta$, the full update up to third order in the weights thus reads:
\begin{equation}
\dot{\boldsymbol{w}} =\frac{1}{M}\sum_{\mu=1}^M \left(c_{2}^{\mu}+c_{4}q^{\mu}\right) C^{\mu}\boldsymbol{w}.
\end{equation}
We can Fourier transform the above equation, exploiting the fact that all the $C^\mu$ are jointly diagonalisable. The previous equation then implies that at the steady state
\begin{equation}
\tilde{w}_{\tau}=0\qquad\text{or}\qquad\sum_{\mu=1}^M\lambda_{\tau}^{\mu}\left(c_{2}^{\mu}+c_{4}q^{\mu}\right)=0,
\end{equation}
where $\tilde{w}_\tau$ are the components of $\boldsymbol{w}$ in the Fourier basis. We thus have a set of $D$ equations with $M+2$ unknown. If follows that $\tilde{w}_{\tau}=0$ for most $\tau$'s.

\subsubsection*{Generic inputs and CP decomposition}
\addcontentsline{toc}{subsubsection}{Generic inputs and CP decomposition}

We decompose the fourth-order moment of the $\mu$th class as $T^{\mu}=T^\mu_{g}+\Delta T^{\mu}$, with $T^\mu_{g}$ and $\Delta T^{\mu}$ the Gaussian component and the fourth-order cumulant, respectively. The full update thus reads:
\begin{equation}
\dot{\boldsymbol{w}} =\frac{1}{M}\sum_{\mu=1}^M\left(c_{2}^{\mu}+c_{4}q^{\mu}\right)C^{\mu}\boldsymbol{w}+\frac{c_{4}}{M}\sum_{\mu=1}^M\Delta T^{\mu}\boldsymbol{w}^{\otimes 3}.
\label{eq:reduced_model_cumulant-methods}
\end{equation}
We employ CANDECOMP/PARAFAC (CP) decomposition \cite{KoldaTensor} in order to find a low-rank approximation of the cumulant tensor $\Delta T$, \emph{i.e.}, a set of $r$ real coefficients $\gamma_a$ and vectors $\boldsymbol{u}_a$ such that
\begin{equation}
    \Delta T_{ijk\ell} \approx \sum_{a=1}^r \gamma_a u_{ia}u_{ja}u_{ka}u_{\ell a}.
\end{equation}
Note that the symmetry of the cumulant tensor implies that the vectors $\boldsymbol{u}_a$ are the same across the 4 modes. For moderate input size $D$, we use the Tensorly package in Python \cite{Tensorly}. For large $D$ (typically for $D>100$), construction and storage of large tensors of higher-order become prohibitive. We thus built upon the framework recently introduced in \cite{KoldaEstimating}, which uses an implicit representation of high-order moment tensors coupled to a gradient-based optimization. We generalized the method in \cite{KoldaEstimating} to deal with the low-rank approximation of cumulant tensors. A detailed description is given in the Supporting Information.

\section*{Acknowledgements}

AI thanks L.F. Abbott for his support and fruitful discussions while at the
  Zuckerman Institute, Columbia University, where research included in this work
  was partly performed. We thank J.~Barbier, M.~Marsili and M.~Refinetti for
  fruitful discussions.


\printbibliography
\appendix
\onecolumn
\numberwithin{equation}{section}
\renewcommand{\thefigure}{S\arabic{figure}}
\setcounter{figure}{0}

\section{Supporting Information}

\subsection{Metastable RF with a plastic second layer} 
\begin{figure}[h]
    \centering
    \includegraphics[width=0.8\textwidth]{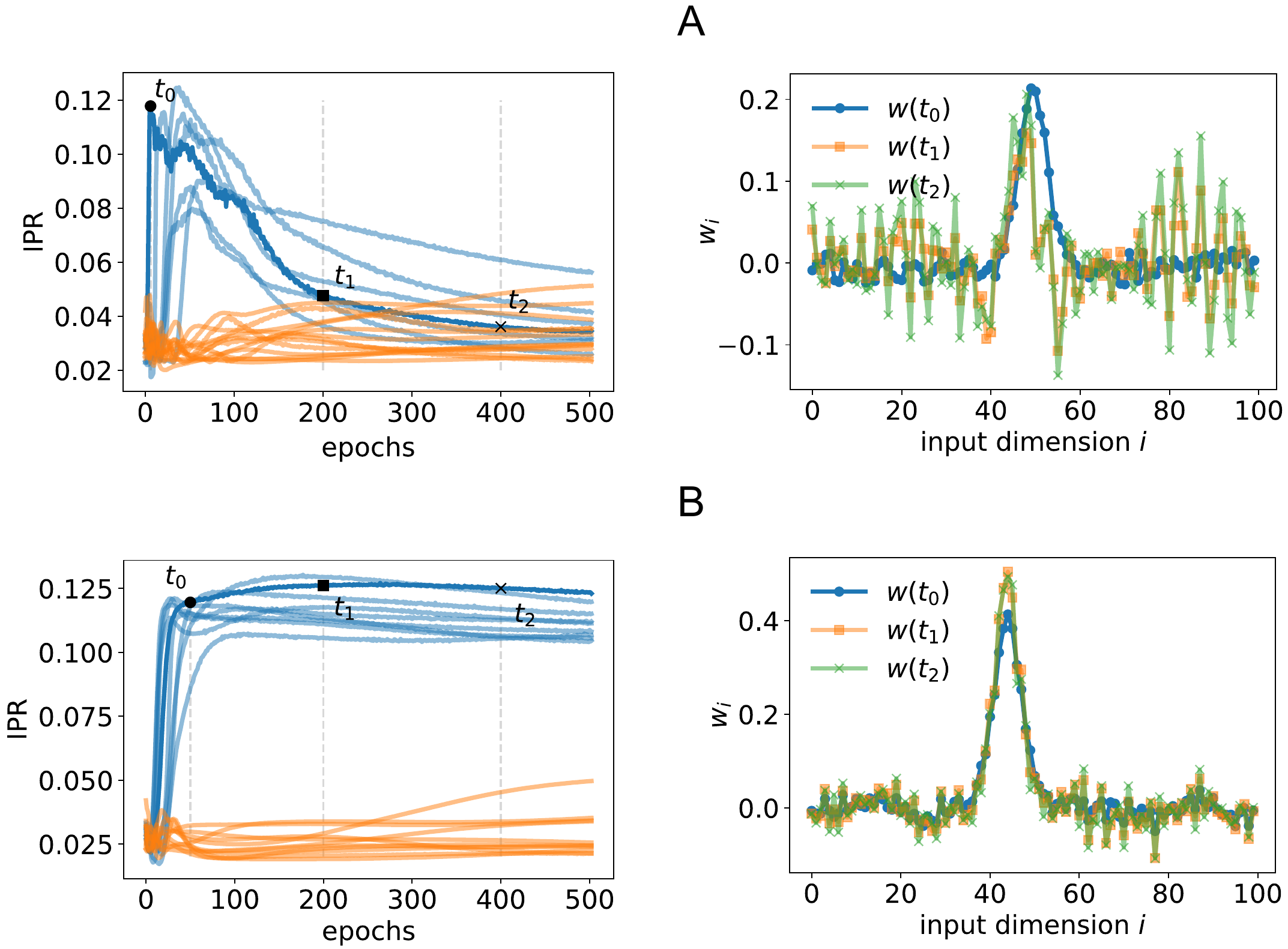}
    \caption{\label{fig:soft_vs_nonsoft} \textbf{Learning dynamics with plastic second layer.} Learning dynamics in a network with $K=20$ hidden neurons solving a 1-dimensional task with $\left(\xi^{-}, \xi^{+}\right)=\left(2,4\right)$ and $D=100$. \textbf{A} and \textbf{B} show the cases where the second layer weights $v_k$ are either trained or fixed at the value $v_k=\frac{1}{\sqrt{K}}$, respectively. Left panels: Inverse Participation Ratio (IPR) of each hidden unit over the course of training epochs. The darker curve indicates the unit shown in the right panels. Right panels: centered receptive fields for a representative unit at three different time-steps during training, marked by the three black symbols in the left plots.
    \emph{Additional parameters}: $g=2$, batch training with $P=\alpha D$ inputs, $\alpha=1000$, learning rate $\eta=0.05$.}
\end{figure}

Our investigation in the main text focused on networks with fixed second layer weights $v_k$. When the second layer is also trained, localised receptive fields emerge at the early stages of learning.
In the left panels of Fig~\ref{fig:soft_vs_nonsoft}, we show the time course of the IPR for $K=20$ hidden units in a network when the second layer is either trained (upper panels) or fixed (lower panels). In the former case, the peaks in the IPR curves signal early localisation in a number of hidden units. As learning unfolds,
initially localised weights acquire an additional oscillatory component, as shown in the overlayed snapshots taken at three different times during training (Fig~\ref{fig:soft_vs_nonsoft}A, right panel). When the second layer is fixed, the localised RF are stable through learning (Fig~\ref{fig:soft_vs_nonsoft}B).

\subsection{Gradient flow with low-rank CP decomposition}
\begin{figure}[tb]
    \centering
    \includegraphics[width=\textwidth]{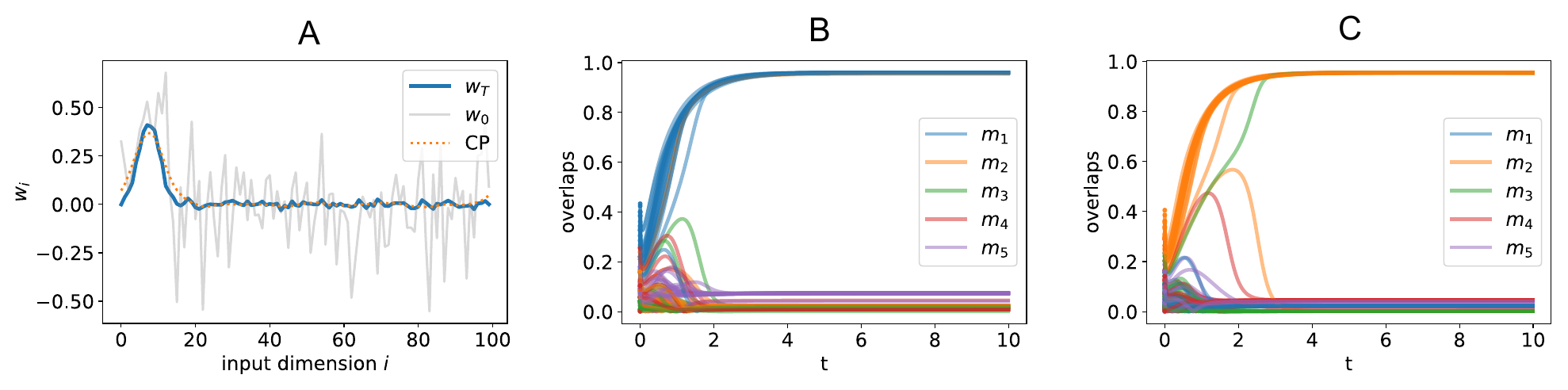}
    \caption{\textbf{Attractor dynamics with Gradient Flow.}
    \textbf{A}: Initial condition $\boldsymbol{w}_0$ (grey line) and steady state $\boldsymbol{w}_T$ (blue line) after $T=1000$ time steps of Gradient Flow dynamics. The red dotted line shows the first CP factor of the cumulant $\Delta T^{2}$.
    \textbf{B}: Time-course of the overlaps $m$'s with the $5$ CP factors $\boldsymbol{u}_r$ for $20$ different $\boldsymbol{w}_0$ initial conditions. The weight vector $\boldsymbol{w}$ is initialized with a small overlap with the first CP factor $\boldsymbol{u}_1$.
    \textbf{C}: Same as \textbf{B} when $\boldsymbol{w}$ is initialized with a small overlap with the second CP factor $\boldsymbol{u}_2$.
    \emph{Additional parameters}: $D=100$, batch training with $P=\alpha D$ inputs, $\alpha=100$, $\left(\xi^{-}, \xi^{+}\right)=\left(0,5\right)$, $\sigma_n$=0.3$, \eta = 0.01$, bias $b$= -1}
    \label{fig:attractor}
\end{figure}

We considered a 1-dimensional task with $\left(\xi^{-}, \xi^{+}\right)=\left(0,5\right)$ and integrated Eq.~5 of the main text using a low-rank reconstruction of the second class cumulant tensor $\Delta T^2$, obtained by CP decomposition. We initialized the weight vector $\boldsymbol{w}$ as a noisy version of the first CP factor $\boldsymbol{u}_1$, namely $\boldsymbol{w}_0=\boldsymbol{u}_1+\sigma_n \boldsymbol{\epsilon}$, with $\boldsymbol{\epsilon}$ an i.i.d.~zero-mean Gaussian vector with unit variance. As shown in \cref{fig:attractor}A, the gradient flow dynamics converges to a spatially localised configuration with a large overlap with $\boldsymbol{u}_1$. We also show in \cref{fig:attractor}B,C the normalized overlaps $m_r=\frac {\boldsymbol{w} \cdot \boldsymbol{u}_r}{\left||\boldsymbol{w}\right||\, \left||\boldsymbol{u}_r\right||}$ with each CP factor $\boldsymbol{u}_r$ when the initial condition $\boldsymbol{w}_0$ has a small initial overlap with $\boldsymbol{u}_1$, $\boldsymbol{u}_2$ respectively.

\subsection{Noise-inducted transitions between receptive fields}
\begin{figure}[tb]
    \centering
    \includegraphics[width=0.5\textwidth]{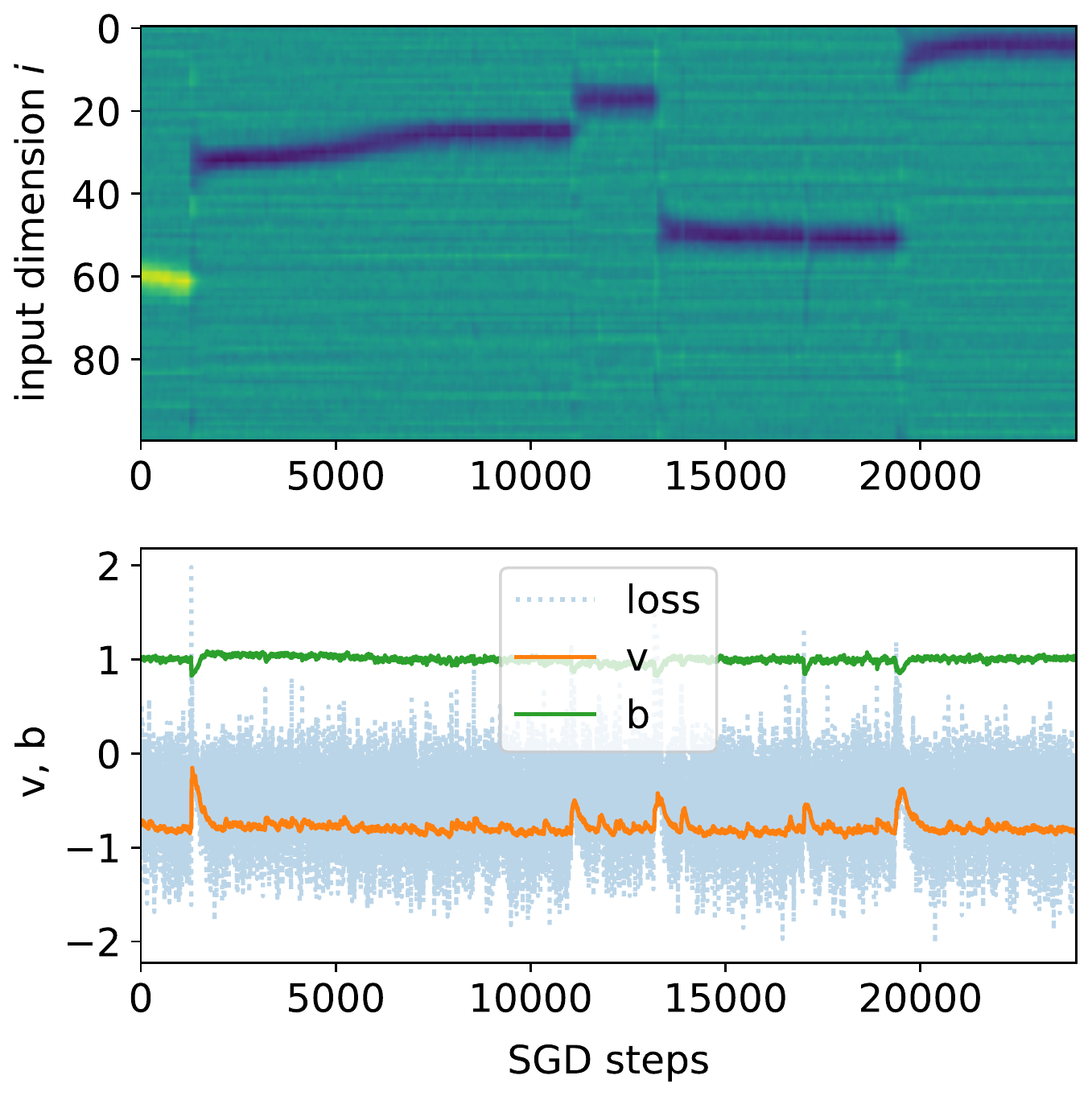}
    \caption{\textbf{Drifting and jumping receptive fields over the course of training with SGD in a network with $K=1$}. Top: Illustration of weight dynamics over a training period of approximately $24000$ gradient steps. Color intensity represents normalized weight value $w_i$. Bottom: time-course of readout weight $v$ and bias $b$ over the same training period. \emph{Additional parameters}: $D=100$, gain $g=3$, batch training with $P=\alpha D$ inputs, $\alpha=20$, $\left(\xi^{-},\xi^{+}\right)=\left(0,4\sqrt{5}\right)$, polynomial activation function $\tilde{\sigma}$, learning rate $\eta=0.01$, batch size=$10$.}
    \label{fig:drift}
\end{figure}

We trained a network with $K=1$ on a 1-dimensional task, using the polynomial activation function $\tilde{\sigma}\left(h\right)=\alpha_1 h-\frac{\alpha_3}{3}h^3$, with $\alpha_1=\sqrt{\nicefrac{2}{\pi}}$ and $\alpha_3=\nicefrac{1}{\sqrt{2\pi}}$.
As explained in the main text, for sufficiently small batch size one observes both drifting periods and transitions between localised fields over the course of learning. Fig~\ref{fig:drift} illustrates the dynamics of the weight vector $\boldsymbol{w}$ when SGD is used with batch of size $10$.

\subsection{Implicit tensor decomposition of higher-order cumulant}
In a recent work~\cite{KoldaEstimating}, Sherman and Kolda introduced a method for computing the CP decomposition of large higher-order moment tensors, exploiting their symmetric structure. We'll focus here on the 4th-order case, which is relevant for our discussion. Given a moment tensor computed from $P$ input patterns $\boldsymbol{x}^\nu$
\begin{equation}
    T_{ijkl}=\frac{1}{P}\sum_{\nu=1}^{P}x_{i}^{\nu}x_{j}^{\nu}x_{k}^{\nu}x_{l}^{\nu}
\end{equation}
one can formalize the CP decomposition in terms of the minimization of the reconstruction loss
\begin{equation}
\tilde{L}=||T-\hat{T}||^{2},
\end{equation}
where $\hat{T}$ is the rank-$r$ tensor
\begin{equation}
\hat{T}_{ijkl}=\sum_{a=1}^{r}\gamma_{a}u_{ia}u_{ja}u_{ka}u_{la}.
\end{equation}
Both the loss function $\tilde{L}$ and its gradient with respect to $\gamma_a$ and $\boldsymbol{u}_a$ can be written solely in terms of the input vectors $\boldsymbol{x}^\nu$, without ever explicitly forming the tensor $T$ (see Eq~$\left(2.6\right)$, $\left(2.7\right)$ and $\left(2.8\right)$ in~\cite{KoldaEstimating}). The best rank-$r$ decomposition can then be obtained using a first-order optimization method of choice.

The generalization to the case of a 4th-order cumulant tensor $\Delta T$ is straightforward. The new loss function reads, up to constant terms:
\begin{equation}
L\left(\left\{ \gamma_{a}\right\} ,\left\{ \boldsymbol{u}_{a}\right\} \right)=\sum_{a,b=1}^{r}\gamma_{a}\gamma_{b}\left(\boldsymbol{u}_{a}\cdot\boldsymbol{u}_{b}\right)^{4}-2\sum_{a=1}^{r}\gamma_{a}\left[\frac{1}{P}\sum_{\nu=1}^{P}\left(\boldsymbol{u}_{a}\cdot \boldsymbol{x}^{\nu}\right)^{4}-3p_{a}^{2}\right],
\label{eq-SM:loss_implicit}
\end{equation}
where $p_{a}=\frac{1}{P}\sum_{\nu}\left(\boldsymbol{x}^{\nu}\cdot\boldsymbol{u}_{a}\right)^{2}$. For the gradients one has:
\begin{align}
\frac{\partial L}{\partial\gamma_{a}}=&-2\left[\frac{1}{P}\sum_{\nu=1}^{P}\left(\boldsymbol{x}^{\mu}\cdot\boldsymbol{u}_{a}\right)^{4}-3p_{a}^{2}-\sum_{b=1}^{r}\gamma_{b}\left(\boldsymbol{u}_{b}\cdot\boldsymbol{u}_{a}\right)^{4}\right]\\
\frac{\partial L}{\partial u_{ia}}=&-8\gamma_{a}\left[\frac{1}{P}\sum_{\nu=1}^{P}\left(\boldsymbol{x}^{\nu}\cdot\boldsymbol{u}_{a}\right)^{3}x_{i}^{\nu}-\frac{3p_{a}}{P}\sum_{\nu=1}^{P}x_{i}^{\nu}\left(\boldsymbol{x}^{\nu}\cdot\boldsymbol{u}_{a}\right)-\sum_{b=1}^{r}\gamma_{b}\left(\boldsymbol{u}_{b}\cdot\boldsymbol{u}_{a}\right)^{3}u_{ib}\right]. \label{eq-SM:grad_implicit}
\end{align}
Following~\cite{KoldaEstimating}, we use Eq~\ref{eq-SM:loss_implicit} and~\ref{eq-SM:grad_implicit} and limited-memory BFGS with bound constraints (L-BFGS-B)~\cite{lbfgs} in a custom version of Tensor Toolbox~\cite{TensorToolbox}.

\subsection{Receptive fields of a two-layer network with a few neurons}

\begin{figure}
    \centering
    \includegraphics[width=0.9\textwidth]{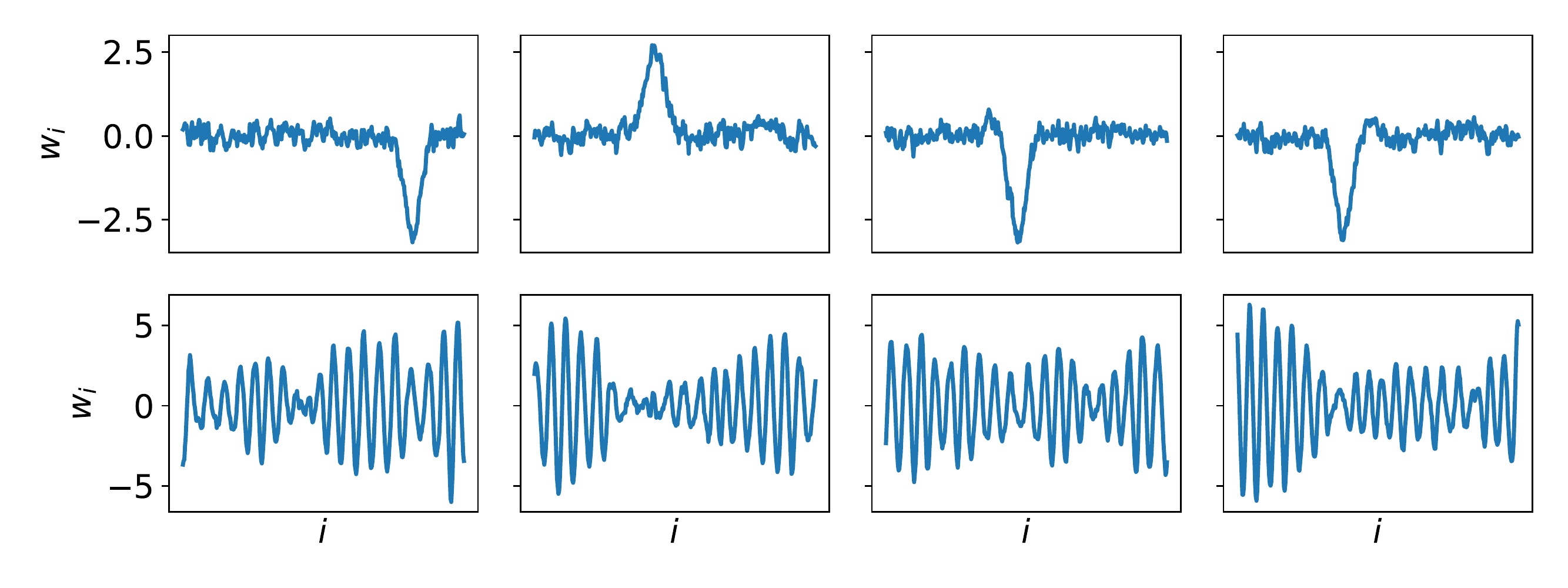}
    \caption{\label{fig:rf_ode} \textbf{Snapshot of receptive fields of a two-layer network in the ODE limit}. Weight vectors of a network trained on a 1-dimensional task with $\left(\xi^{-}, \xi^{+}\right)=\left(8,16\right)$.
    \emph{Additional parameters}: sigmoidal activation function, $v_k=\nicefrac{1}{K}, D=400, K=8, g=3$, online learning, learning rate $\eta=0.2$, weight decay $\ell_2=0.0001$.}
\end{figure}

In \cref{fig:rf_ode}, we plot a snapshot of the receptive fields of a two-layer sigmoidal network trained to discriminate two NLGP processes with correlation lengths $\left(\xi^{-}, \xi^{+}\right)=\left(8,16\right)$. The inputs are one-dimensional with size $D=400$, while the network only has $K=8$ neurons. The network is hence in a limit where a precise description of its dynamics is possible when it is trained on Gaussian inputs~\cite{refinetti2021classifying}. Networks with this architecture develop localised receptive fields (top row), as well as neurons with oscillatory receptive fields, similar to networks whose width is comparable to their input size.

\subsection{Receptive fields of a two-layer ReLU network trained with cross-entropy}%
\label{sec:relu-cross-entropy}

\begin{figure}
    \centering
    \includegraphics[width=0.9\textwidth]{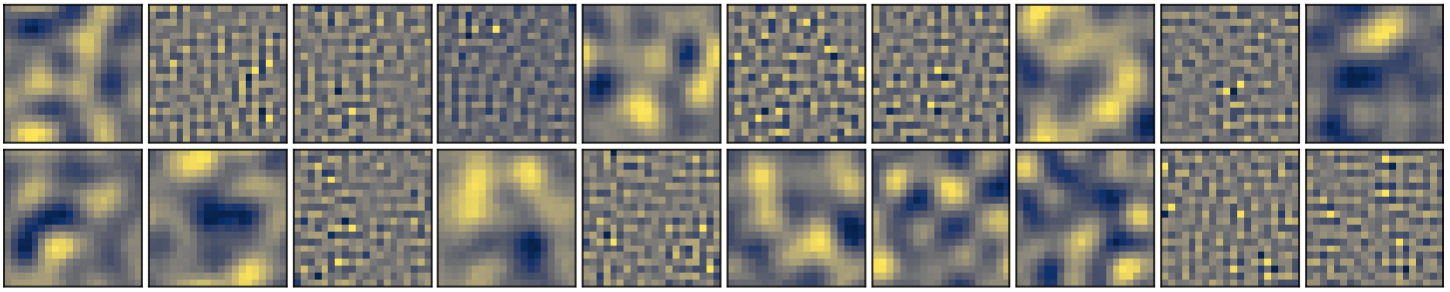}\\[1em]
    \includegraphics[width=0.9\textwidth]{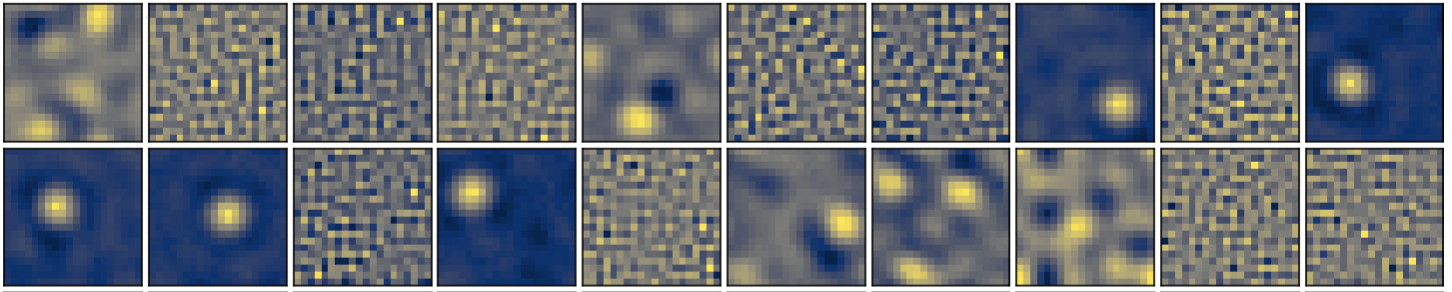}
    \caption{\label{fig:rf_relu} \textbf{Snapshot of receptive fields of a
        two-layer ReLU network trained using the cross-entropy
        loss}. Weight vectors of a network trained in the same setup as Fig.~1,
      except that the activation function is ReLU, $\sigma=\max(0, x)$, and the
      loss is the cross-entropy loss. The top two rows show 20 neurons taken
      from a network trained to discriminate 2-dimensional Gaussian processes
      with $\left(\xi^{-}, \xi^{+}\right)=\left(1, 2\right)$.  \emph{Additional
        parameters}: ReLU activation function,
      $v_k=\nicefrac{1}{K}, L=20, K=100, g=3$, online learning, learning rate
      $\eta=0.2$, weight decay $\ell_2=0.0001$, no bias.}
\end{figure}

We also verified that changing the activation function and the loss function
does not fundamentally alter our results. In \cref{fig:rf_relu}, we plot a
snapshot of the receptive fields of a two-layer network with ReLU activation
function $\sigma(x)=\max(0, x)$, which is especially popular in today's deep learning
architectures. We trained the networks on the binary classification tasks using
the cross-entropy loss function. We see that networks trained on the Gaussian
process again converge to oscillatory receptive fields (top half). Neurons in
the network trained on the non-linear Gaussian process instead develop three
kinds of receptive fields: one with high-frequency oscillations, one with
low-frequency oscillations, and finally there is also a third group of neurons
which shows localised receptive fields.

\end{document}